\documentclass[prb,final,twocolumn,showpacs,preprintnumbers,amsmath,amssymb]{revtex4}

\usepackage[]{graphicx}

\begin{document}

\title{{\textbf{Spin Susceptibility of  Ga-Stabilized $\delta$-Pu Probed by
$^{69}$Ga NMR}}}

\date{\today}
\author{ Yu. Piskunov, K. Mikhalev, A. Gerashenko, A. Pogudin, V.Ogloblichev, S. Verkhovskii, A.
Tankeyev, and V. Arkhipov} \affiliation{Institute  of Metal
Physics, Ural Branch of Russian Academy of Sciences, Ekaterinburg
GSP-170, Russia}
\author{{Yu. Zouev and S. Lekomtsev }}
\affiliation{Russian Federal Nuclear Center - Institute of
Technical Physics, PO Box 245, 456770, Snezhinsk, Russia}

\date{\today}

\begin{abstract}

Spin susceptibility of stabilized $\delta$ phase in the Pu-Ga
alloy is studied by measuring $^{69,71}$Ga NMR spectra and nuclear
spin-lattice relaxation rate $^{69}T_{1}^{-1}$ in the temperature
range 5 - 350 K. The shift ($^{69}K$) of the $^{69,71}$Ga NMR line
and $^{69}T_{1}^{-1}$ are controlled correspondingly by the static
and the fluctuating in time parts of local magnetic field arisen
at nonmagnetic gallium due to transferred hyperfine coupling with
the nearest $f$ electron environment of the more magnetic Pu.  The
nonmonotonic with a maximum around 150 K behavior of
$^{69}K(T)\propto\chi_{s,5f}(T)$  is attributed to the
peculiarities in temperature dependence of  the $f$ electron spin
susceptibility $\chi_{s,5f}(T)$  in  $\delta$ phase of plutonium.
The temperature reversibility being observed in $^{69}K(T)$ data
provides strong evidence for an electronic instability developed
with $T$  in $f$ electron bands near the Fermi energy and
accompanied with a pseudogap-like decrease of $\chi_{s,5f}(T)$  at
$T<150$ K. The NMR data at high temperature are in favor of the
mainly localized character of $5f$ electrons in $\delta$ phase of
the alloy with characteristic spin-fluctuation energy $\Gamma(T)
\propto T^{0.35(5)}$, which is close to $\Gamma(T) \propto
T^{0.5}$ predicted by Cox et al. [J. Appl. Phys. 57, 3166 (1985)]
for $3D$ Kondo-system above $T_{Kondo}$. The dynamic spin
correlations of $5f$ electrons become essential to consider for
$^{69}T_{1}^{-1}(T)$ only at $T<100$ K. However, no NMR evidences
favoring formation of the static magnetic order in $\delta$-Pu
were revealed down to 5K .
\end{abstract}


\maketitle

\smallskip

\section{INTRODUCTION}

Intriguing electronic properties of plutonium and its alloys are
studied intensively over many years bearing an exceptional
position of these materials both in the modern nuclear energy
technologies  and the fundamental  physics of actinides
\cite{Actinides,PlutoniumHandbook}. Metallic plutonium lies in the
actinide series at the boundary between neptunium with itinerant
$5f$ electrons and americium, demonstrating in its magnetic
properties the localized character of $5f$ electrons. The rich
phase diagram of plutonium \cite{Hecker_LAS26,Hecker_JOM55}
presents five polymorphous transitions. The unique structural,
transport and magnetic properties of Pu are determined by the
degree of itinerancy for $5f$ electrons in each of the Pu
allotropes.  Many efforts are undertaken, at present, to elucidate
the ground state of the $f$-electron system  in  $\delta$-Pu and
in stabilized $\delta$ phase Pu-Ga, Pu-Al alloys. Under
theoretical consideration of the equilibrium macroscopic
properties there is assumed either partially
\cite{Eriksson_JAC287} or totally
\cite{Savrasov_PRL84,Kutepov_JPCM15}  localized character of $5f$
electrons, thus fixing a presence of static spin correlations in
the ground state of $\delta$ phase. At the same time, all of
experimental studies regarding to the magnetic state of $\delta$
phase result in dynamic $f$ spin correlations, which development
with temperature does not transform into static magnetic order at
low $T$. The large electronic contribution to specific heat at low
temperature\cite{PlutoniumHandbook,Lashley_PRL91} and the
abnormally low electrical conductivity in  $\delta$ Pu-Ga alloys
\cite{PlutoniumHandbook,Meot-Reymond_JAC232} are known as inherent
electronic properties of the heavy fermions system, where $f$
electrons are in the highly localized state. A growth on-cooling
of the electrical resistivity  in $\delta$-Pu with the following
rather broad maximum of $\rho(T)$ below 200 K is associated with
magnetic scattering of carriers at spin fluctuations of $f$
electrons\cite{Meot-Reymond_JAC232}. A such contribution to
$\rho(T)$ should be  substantially suppressed in the coherent
Fermi-liquid state formed far below $T_{K} \sim 200$ K in
stabilized $\delta$-Pu alloys \cite{Meot-Reymond_JAC232}. The
magnetic susceptibility $\chi(T)$ of the  $\delta$-Pu alloys
Pu$_{0.94}$Ga$_{0.06}$ and Pu$_{0.94}$Al$_{0.06}$
\cite{Meot-Reymond_JAC232} follows modified Curie-Weiss law at
high temperature, favoring the localized state of $f$ electrons
with $\mu_{eff} = 1.2$\quad$\mu_{B}$. A gentle maximum of
$\chi(T)$ near 150 K is explained again  by a Kondo-type effect
which promotes a nonmagnetic ground state. The realistic treatment
of the intra-atomic Coulomb correlations performed for $\delta$-Pu
in Ref.~\onlinecite{Savrasov_PRL84} predicts the structure of $f$
band, implying $5f^{5}$-like atomic configurations of Pu with
$L=5$, $S=5/2$ and $J=5/2$. This feature leads to suppression of
the spin magnetism in $\delta$-Pu.  The suggested structure of $f$
bands is essential for understanding of the weakly temperature
dependent susceptibility which is observed in the stabilized
$\delta$-Pu alloys \cite{Meot-Reymond_JAC232}.

According to the equilibrium Pu-Ga phase diagram
\cite{Hecker_LAS26,Hecker_JOM55}  the stabilized $\delta$ phase is
a metastable structural state of alloys at ambient pressure below
370 K. Indeed, under cooling below 150 K  the partial
$\delta-\alpha^{'}$ martensitic transformation was observed with
characteristic thermal hysteresis of the reverse
$\alpha^{'}-\delta$ transformation on subsequent heating of the
alloy above the room temperature. The low-temperature structural
instability of the Pu-Ga alloys stabilized in $\delta$ phase at
room $T$ gives rise to additional complications in an analysis of
such macro properties as specific heat \cite{Lashley_PRL91},
electrical resistivity \cite{Arko_PRB5} and magnetic
susceptibility \cite{Meot-Reymond_JAC232}. The more reliable data
regarding to the spin fluctuation regime of $5f$ electrons in
$\delta$ phase of Pu are provided by the locally sensitive neutron
scattering  and NMR experiments. The results presented under such
kind local studies can be considered as a key moment to clarify
details of thermally- or pressure-induced structural instability
both of the pure plutonium metal and  the Pu-based alloys.
Unfortunately, the early NMR studies performed in 60-s present a
rather fragmentary raw of data regarding to the $^{27}$Al NMR in
Pu$_{0.95}$Al$_{0.05}$ \cite{Fradin_IJM1} or unsuccessful efforts
in detection of the $^{239}$Pu NMR signal \cite{Carter_MS20}
without any subsequent  analysis of the results obtained.

In this paper we report on $^{69,71}$Ga NMR being applied for the
first time to trace the temperature dependence of a spin
susceptibility $\chi_{s,5f}(T)$ in the range 5 - 350 K for
Ga-stabilized $\delta$-Pu. The shift ($^{69}K$) of the
$^{69,71}$Ga NMR line (transition ($-1/2\longleftrightarrow
+1/2$)) and $^{69}T_{1}^{-1}$ are controlled correspondingly by
the static and fluctuating in time parts of local magnetic field
arisen at nonmagnetic gallium due to transferred hyperfine
coupling with the nearest $f$ electron environment of the more
magnetic Pu. The reversible with $T$ nonmonotonic behavior of
$^{69}K(T) \propto \chi_{s,5f}(T)$ provides strong evidence for an
electronic instability developed with $T$ in $f$ electron bands of
$\delta$-Pu  similar to the observed by NMR in concentrated
Kondo-systems of Ce \cite{Sarrao_PRB59,Curro03}. A such kind of
instability may be considered as one of driving forces for the
martensitic $\delta-\alpha^{'}$ transformation observed recently
in Pu$_{1-x}$Ga$_{x}$ $(x< 0.04)$ alloys below 150 K. The NMR data
at $T > 100$ K are in favor of the  mainly localized character of
$5f$ electrons in $\delta$ phase of the alloy with characteristic
spin-fluctuation energy $\Gamma(T) \propto T^{0.35(5)}$, which is
close to $\Gamma(T) \propto T^{0.5}$ predicted for $3D$
Kondo-system above $T_{K}$ by Cox et al.\cite{Cox85}. The dynamic
spin correlations of $5f$ electrons become essential to consider
for $^{69}T_{1}^{-1}(T)$ only at $T <100$ K.

\section{EXPERIMENT}

The polycrystalline sample of  $\delta$-Pu alloy having a nominal
Ga concentration of 1.5 wt \% was prepared as a thin plate
($10\times3.5\times0.2$ mm). Surface of the plate was mechanically
polished with subsequent electrochemical etching to remove any
impurities and the oxide layer induced during preparation of the
sample by machining from ingot of alloy. Immediately after
cleaning the plate was housed under a dry argon atmosphere in a
cylinder glass container, which finally was soldered from both
sides for safety.

NMR measurements were performed in magnetic field
$H_{0}=\omega_{0}/\gamma_{Ga}=94$\quad kOe using a phase-coherent
pulse spectrometer supplied with a quadrature receiver. The
$^{69,71}$Ga ($I = 3/2$) spectrum of central transition
($-1/2\longleftrightarrow+1/2$) was obtained by Fourier
transformation of the second half of the spin-echo signal followed
the $(\pi/2)_{x}-t_{del}-(\pi)_{x}$ pulse sequence. The frequency
band excited by the $\pi$  pulse was about 100 kHz and the whole
spectrum including all of transitions was measured by summation of
subsequent Fourier-signals of an echo accumulated at different
equidistant operating frequencies. Below 40 K where self-heating
effects should be taken into account we  check additionally the
temperature of sample with accuracy ($\Delta T/T\sim0.02$) by
comparing the Ga NMR line intensity  measured at the different
$T$.
\begin{figure}[htbp]
\centerline{\includegraphics[width=0.95\hsize]{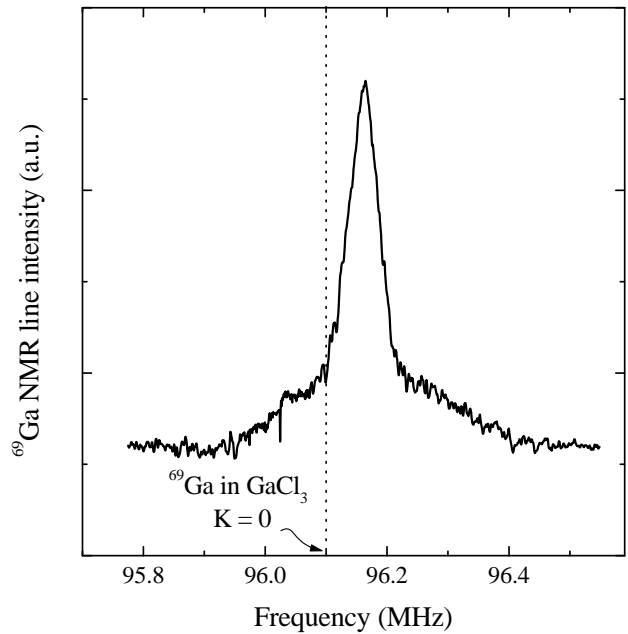}}
\caption{$^{69}$Ga NMR spectrum in Ga-stabilized $\delta$-Pu  at
20 K. The dotted line shows the NMR line position $^{69}\nu_{0}  =
96.103$ MHz ($H_{0} = 94$ kOe) in GaCl$_{3}$ used as a standard
reference.} \label{fig1}
\end{figure}

\bigskip

\section{ RESULTS AND DISCUSSION}

The $^{69}$Ga NMR spectrum for Pu-Ga alloy shown in
Fig.~\ref{fig1} consists of a single nearly gaussian central line
($ -1/2 \longleftrightarrow +1/2$) transition and a broad pedestal
contributed by the satellite ($\pm1/2 \longleftrightarrow \pm3/2$)
transitions. A similar pattern of the spectrum is observed for a
powder of imperfect cubic crystals \cite{Abragam}. Thus local
distortions of the cubic symmetry take place for charge
environment of the Ga nuclei possessing by rather large electric
quadrupole moment $^{69}Q = 0.168\times10^{-24}$ cm$^{-2}$ and
$^{71}Q = 0.106\times10^{-24}$ cm$^{-2}$. Distribution of the
electric field gradient arisen due to cubic imperfections results
in substantially broad pedestal of satellite lines with a width
$(\Delta \nu_{Q})\propto e^{2}QV_{zz}$ . As found, below 100 K the
pedestal width $^{69}(\Delta \nu_{Q})\sim 0.5$ MHz stays unchanged
thus indicating that cubic imperfections near the NMR probe - Ga
are stable over the temperature range 5 -100 K for $\delta$ phase
of Pu. Quadrupole broadening effects on the width of the central
line $\Delta \nu_{Q}(-1/2 \longleftrightarrow +1/2) \propto \Delta
\nu_{Q}^{2}/\nu_{0}$ is greatly reduced in high magnetic field and
its contribution to the width of $^{69}$Ga line does not exceed 2
kHz.
\begin{figure}[htbp]
\centerline{\includegraphics[width=0.95\hsize]{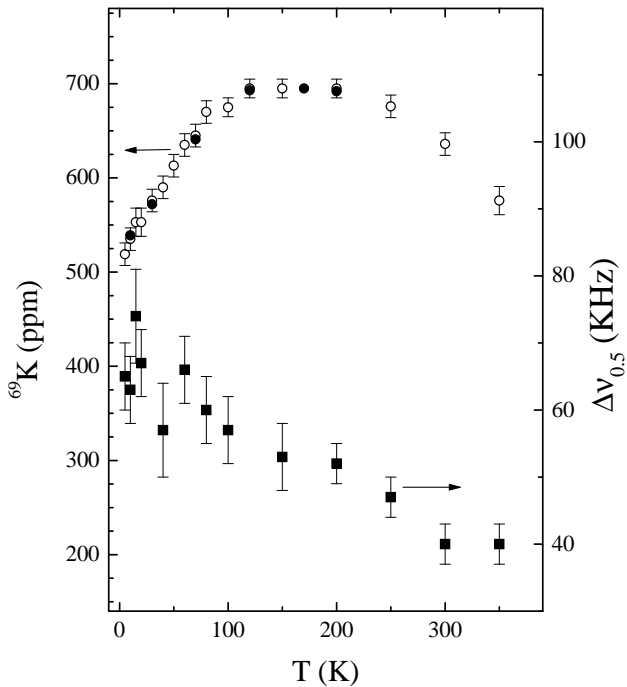}}
\caption{The shift ($^{69}K$) and the width ($\Delta\nu_{0.5}$)
taken at a half of  height of the $^{69,71}$Ga NMR line
(transition ($-1/2\longleftrightarrow+1/2)$)  in stabilized
$\delta$ phase of Pu-Ga  alloy versus $T$. The $^{69}K(T)$ data
marked by the open circles ($\circ$) are obtained on heating the
sample and solid circles ($\bullet$) correspond to data obtained
during further thermal recycling of the sample over temperature
range 5 - 350 K to testify reversibility with temperature of the
$K(T)$ data.} \label{fig2}
\end{figure}
By comparing spectra measured for two Ga isotopes, which nuclear
gyromagnetic ratio differentiate in $^{71}\gamma/^{69}\gamma
\approx 1.26$ times, we sure that the width of central line
($\Delta \nu_{0.5}$) is magnetic in origin. Its magnitude is
determined by a distribution of local magnetic field created at
Ga-sites by the nearest environment of the more magnetic Pu.
Magnetic broadening of central line exceeds more than three times
the maximally possible contribution of classic dipolar field,
assuming for such estimate the ferromagnetic spin order among
twelve neighboring Pu with $\mu(5f)\sim5\mu_{B}$ per Pu. The main
contribution to the width of central line is expected due to the
effect of spin polarization of conducting electrons  through RKKI
coupling with $5f$ electrons of Pu.

 A gradual increase of $\Delta \nu_{0.5}$ with decrease of $T$
evidences for the growth of the short-wave contributions to spin
susceptibility of $\delta$ phase in alloy. However, a typical
abnormal steep broadening of the NMR line  favoring formation any
kind of static magnetic order in the $\delta$-Pu was not observed
down to 5 K. That is in line with the results of macroscopic
magnetization $M(H)$ measurements performed up to 5 T at $T = 1.8$
K for the very same sample of Pu-Ga  alloy.
\begin{figure}[htbp]
\centerline{\includegraphics[width=0.95\hsize]{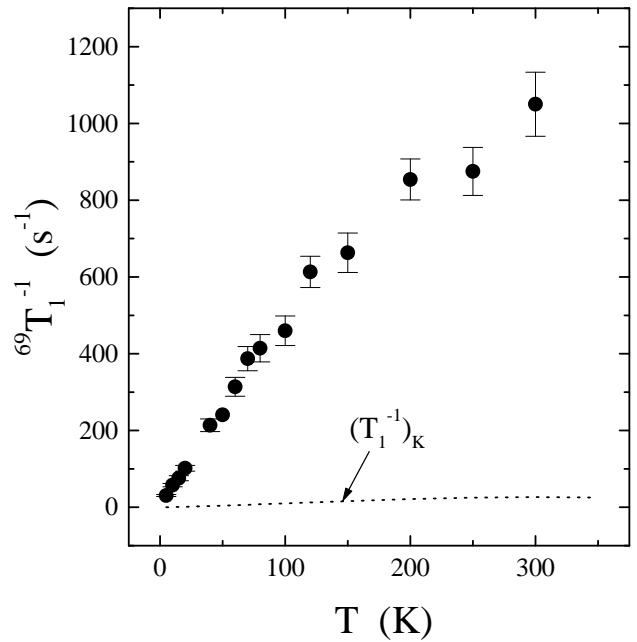}}
\caption{Nuclear spin-lattice relaxation rate $T_{1}^{-1}$ of
$^{69}$Ga in Ga-stabilized $\delta$-Pu  versus $T$. The dotted
line shows the contribution of conducting electrons
$(T_{1}^{-1})_{K}$ estimated from Korringa relation (Eq.
(\ref{Korringa}), see text).} \label{fig3}
\end{figure}

In the alloy the peak of central line ($\nu_{p}$) is shifted to
the higher frequency region with respect to NMR line position
$^{69}\nu_{0}=96.103$ MHz in GaCl$_{3}$ used as a standard
reference. Symmetric shape of the central line implies that the
$^{69,71}$Ga NMR shift $K = (\nu_{p}-\nu_{0})/\nu_{0}$ is an
isotropic quantity inherent for crystals of cubic point symmetry.
The main part of $K(T)$ data marked by the open circles ($\circ$)
in Fig.~\ref{fig2} were collected on heating the sample between
subsequent measurements of the $^{69}$Ga NMR spectrum. The $K(T)$
data shown by solid circles ($\bullet$) were obtained further
during a thermal recycling of the sample over temperature range 5
- 350 K, to testify reversibility with temperature of the $K(T)$
data row. An absence of any hysteresis in $K(T)$ data no any new
line would be appeared  in  replicates of the NMR measurements
on-cooling evidence strongly that we have deal with NMR line of Ga
sited in stabilized $\delta$ phase (the FCC structure
\cite{Hecker_LAS26,Hecker_JOM55} ) of the Pu-Ga alloy.

The Pu alloys stabilized by Al\cite{Lashley_PRL91} or
Ga\cite{Meot-Reymond_JAC232} in $\delta$ phase remind in its
electronic properties the heavy fermion behavior and keeping in
mind a such reminiscence we consider the total NMR shift for
nonmagnetic Ga in $\delta$ phase as the Knight shift $K_{0}$
arisen due hyperfine coupling with carriers in the conduction
band. An effect of the additional uniform spin polarization  of
conducting electrons through RKKI like interaction with much less
itinerant $5f$ electrons is taken into account as an additive term
$K_{f}$:
\begin{equation}
K(T) = K_{0} +
K_{f}(T)=K_{0}+\frac{12H_{f}}{\mu_{B}N_{A}}\chi_{s,5f}(T),
\label{shift}
\end{equation}
here $H_{f}$ - an effective hyperfine magnetic field created at
the Ga nuclei by the unpaired electron spin from  $5f$ shell of
one of twelve neighboring Pu in FCC structure of $\delta$ phase.
$N_{A}$ - Avogadro's number and $\chi_{s,5f}$ - the molar spin
susceptibility of $f$ electrons for stabilized $\delta$ phase of
the Pu alloy.

The nonmonotonic temperature dependence of $^{69}K(T)$ with a
maximum near 150 K pays on itself our attention. A similar maximum
in $\chi(T)$ was  revealed recently in measurements of static
magnetic susceptibility for Pu$_{0.94}$Ga$_{0.06}$
alloy\cite{Meot-Reymond_JAC232} and it was attributed to the Kondo
anomaly $f$ electrons  in spin susceptibility $\chi_{s}(q=0)$ of
the alloy. The great difference in "modulation index" of $K(T)$
(present work) and $\chi(T)$ (Ref.
\onlinecite{Meot-Reymond_JAC232}) for the stabilized $\delta$
phase Pu alloys with the very same concentration of Ga is caused
apparently by the $T$-independent Van Vleck contribution
dominating  in $\chi(T)$ \cite{Savrasov_PRL84}. The temperature
dependence of a  spin contribution to $\chi(T)$  can be masked by
any kind of spurious contributions in macroscopic measurements of
$\chi$. On other side, a puzzling similarity in temperature
dependences of $K(T)$ and $\chi(T)$ lets us to put in further
estimates that  NMR shift of Ga is almost  determined by the
second term in Eq. (\ref{shift}) which is proportional to
$\chi_{s,5f}$. Then using a linear fit of  the $^{69}K(T)$ vs
$\chi(T)$ \cite{Meot-Reymond_JAC232} plot with $T$ as a parameter,
we get an estimate of the effective hyperfine magnetic field
$H_{f} = 2.8(6)$\quad kOe/$\mu_{B}$.

Thus for the first time the temperature dependence of shift of the
$^{69}$Ga NMR shift presented in this paper  provides strong
evidence for an electronic instability developed with temperature
in f electron bands and accompanied below 150 K with a
pseudogap-like decrease of spin susceptibility in stabilized
$\delta$  phase of the Pu-Ga alloy. Apparently, the similar
changes in electronic system  at low T take place in a $\delta$
phase of Pu stabilized by Al, if to take into account the tendency
of decrease with temperature of the $^{27}$Al NMR shift observed
earlier in Pu$_{0.95}$Al$_{0.05}$ \cite{ Fradin_IJM1,Carter_MS20}.

Further the temperature dependence of spin-lattice relaxation rate
of $^{69}$Ga nuclei is considered to elucidate peculiarities in
the low frequency spin dynamic of f electrons in Ga-stabilized
$\delta$-Pu. Corresponding characteristic time interval $T_{1}$
for relaxation of longitudinal nuclear magnetization $M_{z}(t)$ to
its equilibrium value $M_{0}$ was measured by an
inversion-recovery technique using the $\pi - t - \pi/2 - \tau-
\pi$  pulse sequence. An intensity of echo formed by central line
of $^{69}$Ga was measured as function of the time interval $t$
while $\tau$ is fixed. Duration of $\pi$  pulse corresponds to
inversion at $t = 0$ of the population between $m_{I} = -1/2$ and
$m_{I} = +1/2$ levels in spin system of $^{69}$Ga ($I=3/2$).
Subsequent recovery in time of ($M_{0}-M_{z}(t)$) was fitted with
an expression representing  two exponents having characteristic
time of $T_{1}$ and $T_{1}/6$ and summed with corresponding
weights of 0.1 and 0.9 \cite{Andrew_PrPSL_78}.
\begin{figure}[htbp]
\centerline{\includegraphics[width=0.95\hsize]{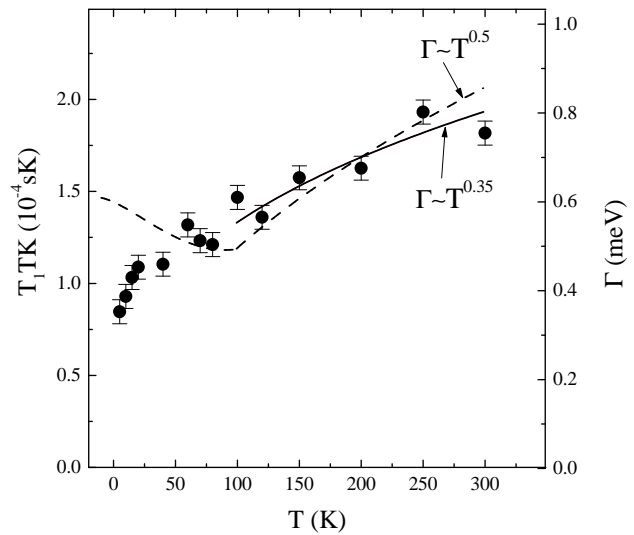}}
\caption{Temperature dependence of the characteristic energy of
$f$ spin fluctuations $\Gamma(T)\propto T_{1}TK$. Solid and doted
lines for $T > 100$ K represent curves $\Gamma(T) \propto
T^{0.35(5)}$ (the best fit for $\Gamma$ data taken above 100 K)
and $\Gamma(T) \propto T^{0.5}$ respectively. At $T<100$ K the
doted curve demonstrates behavior of  $\Gamma(T)$ predicted by Cox
et al. in Ref. \onlinecite{Cox85} Kondo system at low temperature
region $T<T_{K}$.} \label{fig4}
\end{figure}

The  $^{69}$Ga spin-lattice relaxation rate data  in stabilized
$\delta$ phase of plutonium are plotted versus $T$ in
Fig.~\ref{fig3}. For nonmagnetic element Ga surrounded in the
alloy with much more magnetic actinide atoms of Pu the nuclear
spin-lattice relaxation rate of Ga is determined by two
contributions:
\begin{equation}
\frac{1}{T_{1}}=\left(\frac{1}{T_{1}}\right)_{K}+\left(\frac{1}{T_{1}}\right)_{f}
\label{sumrates}
\end{equation}
The first contribution ($T_{1}^{-1})_{K}$ is caused by
Fermi-contact interaction of nuclear spin with conducting
electrons and its crude estimate can be obtained from the well
known Korringa relation, taken in the form accepted for the free
electron gas model  of metal:
\begin{equation}
\left(\frac{1}{T_{1}}\right)_{K}=\frac{h\gamma_{n}^{2}k_{B}}{2\mu_{B}}TK_{s}^{2},
\label{Korringa}
\end{equation}
where $K_{s}$ is a Knight shift. The second term in
Eq.(\ref{sumrates}) determines the channel of nuclear spin
relaxation caused by time-dependent fluctuations of local field
determined by spin of $f$ electrons through the RKKI interaction.
Following T. Moriya \cite{Moriya_JPSJ18} the contribution
$(T_{1}^{-1})_{f}$ is proportional to the q-weighted imaginary
part of the dynamic spin susceptibility
$\chi_{f}(q,\omega\approx\omega_{0})$ of $f$ electrons:
\begin{equation}
\left(\frac{1}{T_{1}}\right)_{f}=\frac{\gamma_{n}^{2}k_{B}T}{2\mu_{B}^{2}}\sum_{q}H_{f}^{2}(q)\frac{\chi_{f}^{''}(q,\omega_{0})}{\omega_{0}}.
\label{fcontr}
\end{equation}
Here $H_{f}(q)$ is a structural form-factor, determining hyperfine
magnetic field at the Ga nuclei from $5f$ electrons of Pu,
$\omega_{0}$ - the nuclear Larmor frequency. If to ignore both the
spin correlations among $f$ electrons and the q-dependence of
$\chi_{f}$ , then the general expression Eq.(\ref{fcontr}) for
$(T_{1}^{-1})_{f}$ is reduced to the following one:
\begin{equation}
\left(\frac{1}{T_{1}}\right)_{f}=\frac{\gamma_{n}^{2}k_{B}TzH_{f}^{2}}{2\mu_{B}^{2}}\chi_{s,5f}(T)\frac{\tau}{2\pi},
\label{fcontr2}
\end{equation}
where $z$ is a number of the nearest neighbor of Pu around
$^{69}$Ga used as NMR and $\chi_{s,5f}(T)$ is a static spin
susceptibility of $f$ electrons. Finally, by taking into account
that for Ga in alloy $K\approx\frac{zH_{f}}{\mu_{B}}\chi_{s,5f}$
we get:
\begin{equation}
T_{1}TK=\frac{2\mu_{B}}{\gamma_{n}^{2}k_{B}H_{f}}\Gamma(T).
\label{gamma}
\end{equation}
Here  $\Gamma\sim\tau^{-1}$ is the characteristic energy of $f$
spin fluctuations, being inverse proportional to  $\tau^{-1}$ -
the rate of spin relaxation for $5f$ electrons of Pu. For
comparison in strength of two terms in Eq.(\ref{sumrates}) we have
drawn with dotted line in Fig.~\ref{fig3}  the temperature
dependence of $(T_{1}^{-1})_{K}$ as estimated from the Korringa
relation in Eq.(\ref{Korringa}). As seen, this contribution is
roughly forty times less of the observed. Thus, one may suggest
that spin-lattice relaxation rate of Ga nuclear spin is fully
controlled in stabilized $\delta$ phase of Pu by fluctuating in
time magnetic fields of $f$ electrons.  One can quantify the
$\Gamma(T)$ behavior for stabilized $\delta$-Pu over the
temperature range under this study by using expression
Eq.(\ref{gamma}) with $H_{f}=$ 2.8\quad kOe/$\mu_{B}$.  The
results obtained for $\Gamma(T)$ are shown by solid circles in
Fig.~\ref{fig4}. Above 100 K the characteristic energy of spin
fluctuations is increased following $\Gamma(T) \propto
T^{0.35(5)}$ (corresponding fitting curve is drawn by solid line
in Fig.~\ref{fig4}). A such behavior of  $\Gamma(T)$ is close to
$\Gamma(T) \propto T^{0.5}$ predicted by Cox et al.\cite{Cox85}
for Kondo-systems above the Kondo temperature $T_{K}$. A
reasonable deviation of the $\Gamma(T)$-data at $T < 100$ K from
the behavior predicted by Cox et al. \cite{Cox85} for the coherent
Fermi-liquid state formed below $T_{K}$ (see dotted curve in
Fig.~\ref{fig4}) may indicate on importance of taking into
consideration for $(T_{1}^{-1})_{f}$ of the antiferromagnetic spin
correlations between $f$-electron of the nearest Pu atoms. In
accordance with Eq.(\ref{fcontr}) one can expect a sizeable growth
of the short-wave ($q\sim Q_{AF}$) contributions to
$(T_{1}^{-1})_{f}$. Here $Q_{AF}$ is a wave vector of the
corresponding spin order, which would arise due to incomplete
suppression of the spin degree of freedom for the $f$ electronin
the framework of conventional Kondo-scenario developed for system
with heavy fermions.
\section{ CONCLUSION}
In summary, we have measured the temperature dependence of the
$^{69,71}$Ga Knight shift and the nuclear spin-lattice relaxation
of $^{69}$Ga in Ga-stabilized $\delta$ phase of the Pu-Ga alloy
with a goal to elucidate the behavior of spin susceptibility
$\chi_{s,5f}(T)$ in $\delta$ phase of the alloy over the
temperature range 5 - 350 K. The reversible with $T$  nonmonotonic
behavior of the Khight shift evidences for the instability of the
electron spectrum near the Fermi energy developed in $\delta$
phase below 150 K and accompanied with a pseudogap-like decrease
of  the uniform ($q\sim0$) part of  spin susceptibility. The
experimental results regarding spin-lattice relaxation rate of
$^{69}$Ga do not contradict to an assumption that $\delta$ phase
stabilized by Ga in its electronic properties can de treated as
the $5f$-concentrated Kondo system with $T_{K}\geq100$ K. However
it should to remark that unambiguous NMR conclusions concerning an
itinerancy of $5f$ electrons  in $\delta$-Pu can be done after NMR
studies at high temperature of the stabilized $\delta$ phase Pu-Ga
alloys. These measurements are now in progress.
\begin{acknowledgments}
The authors wish to thank the Director of Institute of Technical
Physics G.N. Rykovanov for supporting of the present work. We are
also grateful to A. P. Pokatashkin and N. A. Kiselev for the
solution of complicated organizational matters, to E. A. Shmakov,
D. V. Taskin,  N. A. Tsepilov and N. A. Aleksashina for samples
preparation. We thank I. L. Svyatov and D. V. Yakovlev for
numerous kind of assistance. Y.P. acknowledges the "Russian
Science Support Foundation" for a financial support. The work was
supported by the Ministry of Sciences and Technology of the
Russian Federation under contract No 25/03.
\end{acknowledgments}

\bibliographystyle{apsrev}
\bibliography{ref}

\end{document}